# Polariton Topological Insulators with Disclinations

**Khalil Sabour[1] and Yaroslav V. Kartashov[2]**

[1]*Moscow Institute of Physics and Technology, Institutsky lane 9, Dolgoprudny, Moscow region, 141701, Russia*
[2]*Institute of Spectroscopy, Russian Academy of Sciences, 108840, Troitsk, Moscow, Russia*

**Abstract** We introduce polariton topological insulator on aperiodic array of microcavity pillars with disclination. This unique dissipative and nonlinear topological system can have discrete rotational symmetry not achievable in structures constructed on periodic arrays that qualitatively affects the structure of topological modes. In such system resonant optical pump can be used for selective excitation of topological states forming on the disclination core in the center, while imposing vorticity on the pump profile allows to excite not only the simplest disclination states, but also vortex-carrying disclination states. Strong repulsive nonlinearity of polaritonic system leads to increasing with pump amplitude tilt of the resonance curves and bistability, with rich possibilities for control of shapes of nonlinear disclination states within topological gap by pump frequency detuning. Stability analysis predicts remarkable robustness of both vortex-free and vortex-carrying nonlinear disclination states in this system. We show that when topological charge of the pump exceeds maximal allowed charge in the disclination array with a given discrete rotational symmetry, the stationary states form that contain vortex clusters in their phase distributions. Our results extend the concept of topological insulators with disclinations to the realm of *nonlinear resonantly driven* systems and show that rich bistability effects in them can be observed even for excited vortex-carrying topological states with different topological charges allowed by discrete rotational symmetry of the system.

## 1. Introduction

Topological insulators (TIs) are novel materials that are insulating in the bulk but allowing propagation of excitations along their edges. Robustness of such edge currents is guaranteed by the nontrivial topology of bulk bands of TI. Initially discovered in electronic systems [1,2], TIs were later studied and found applications in various areas of physics including acoustics [3], cold atoms and physics of matter waves [4,5], photonics [6–9] and physics of polariton condensates [10, 11]. In many cases, the observation of topological edge states followed development of new technologies for the fabrication or induction of structures with topologically nontrivial bands, as it happened, e.g., for polaritons in microcavities [10,12,13]. Nontrivial band topology, manifesting itself in the formation of topological states at the edges of the material, usually emerges upon breakup of certain symmetries of the system, such as time-reversal or inversion symmetry, it can also stem from deformations of the unit cells or temporal modulations of system parameters [7,9]. Thus, in the intriguing class of higher-order topological insulators (HOTIs) topological phases can be realized by appropriately shifting sites within the unit cell of a structure that affects intra- and intercell couplings. In contrast to conventional $d$ -dimensional insulators that support only $(d-1)$ -dimensional edge states, the HOTI of the $n$ -th order $(n>1)$ created on a $d$ -dimensional lattice can support in topological phase the boundary states with effective dimensionality $(d-n)$ , i.e. in these intriguing structures the effective dimensionality of topological boundary modes can be more than one dimension below that of the bulk, see [14-16]. For instance, second-order two-dimensional HOTI can support effectively zero-dimensional corner states in addition to one-dimensional edge states, as observed in various photonic [17-22] and acoustic [23-26] systems.

Very recently it has been discovered that HOTIs can be created not only on the basis of periodic lattices, but also using aperiodic lattices with disclinations – crystallographic defects that globally modify the lattice structure, while creating materials with new discrete rotational symmetries and unconventional transport properties, not accessible in periodic HOTIs [27-30]. These global defects can trap fractional spectral charges (linked to the local density of states [14,28,29]) and host localized topological states at the disclination core in addition to topological states at the outer edges and corners [31-33]. Recently the concept of disclination states has been extended to discrete non-Hermitian lattices with linear gain and losses on individual sites that cannot possess parity-time symmetry due to structure of the lattice and on this reason always has complex linear spectrum with multiple coexisting bulk and topological modes with comparable growth rates [34]. HOTIs with disclinations have been shown to support nonlinear and vortical disclination states [35-38] but primarily within conservative systems. Nonlinearity in such systems introduces a convenient knob for control of localization, structure and evolution dynamics of topological states. A notable exception is a laser cavity with a photonic disclination reported in [39], where gain saturation provides a nonlinear mechanism leading to stable lasing. However, the formation and control of nonlinear topological vortex states in *resonantly driven*, dissipative HOTIs with disclinations built on aperiodic lattices remain unexplored. Here, we demonstrate how such nonlinear disclination modes can be robustly excited and controlled via optical pumping in polaritonic system.

In this work, we propose a nonlinear dissipative HOTI on aperiodic array of microcavity pillars with disclination, where polariton condensation can occur. This unique topological system can feature discrete rotational symmetry that is unattainable in periodic arrays, fundamentally altering the structure of its topological modes, and, at the same time, it is characterized by a very strong polariton-polariton interactions. Notice that previously polariton microcavity arrays were suggested as a platform for realization of various nonlinear Chern insulators supporting unidirectional edge states and solitons [10,40-52]. Polaritonic HOTIs on periodic lattices have been suggested in [53-56], but first experiments on corner states in two-dimensional Su-Schrieffer-Heeger lattices were reported only recently with the development of perovskite microcavity systems [57,58], see also [59,60].

We show that resonant optical pump [61-63] in our system, that counteracts intrinsic losses, enables selective excitation and control of shapes of topological in-gap states localized at the disclination core, including vortex-carrying ones (for structured pump with imposed vorticity). Strong repulsive nonlinearity in this system results in a nonlinear tilt of the resonance curves and bistability, offering rich control over the shapes of nonlinear disclination states within the topological gap via pump frequency detuning. Stability analysis predicts remarkable robustness of both vortex-free and vortex-carrying dissipative nonlinear disclination states.

## 2. The model and linear spectrum of the system

We consider the evolution of polariton condensate in the array of microcavity pillars with disclination, governed by the dimensionless Gross-Pitaevskii equation for the polariton wavefunction $\psi$ [11]:

$$i\frac{\partial\psi}{\partial t} = -\frac{1}{2}\Delta\psi - \mathcal{R}(x,y)\psi + |\psi|^2\psi - i\alpha\psi + \mathcal{P}(x,y)e^{-i\varepsilon t}. \quad (1)$$

Here $\Delta = \partial^2/\partial x^2 + \partial^2/\partial y^2$; the coordinates $(x,y) = \mathbf{r}$ are scaled to the characteristic length $\ell$ that allows to introduce the characteristic energy scale $\varepsilon_0 = \hbar^2 / m_{\mathrm{eff}}\ell^2$, where $m_{\mathrm{eff}}$ is the effective polariton mass for lower polariton branch; the evolution time $t$ is normalized to $\hbar\varepsilon_0^{-1}$; $\varepsilon$ is the pump frequency detuning from the bottom of the lower polariton branch normalized to $\hbar^{-1}\varepsilon_0$; $\alpha = \hbar / \tau\varepsilon_0$ is the loss coefficient, where $\tau$ is the polariton lifetime; $\Psi = (\varepsilon_0 / g)^{1/2}\psi$ is the dimensional wavefunction, $g$ is the interaction constant; the function $\mathcal{R}(x,y) = p\sum_{m,n} e^{[-(x-x_{m,n})^2-(y-y_{m,n})^2]/w_0^2}$ models the potential energy landscape $(-\mathcal{R})$ in the array of pillars, scaled to $\varepsilon_0$ and composed from wells with half-width $w_0$ and depth $p$, located in the points $x_{m,n}, y_{m,n}$. Nonlinear term accounts for repulsion between polaritons; the function $\mathcal{P}(x,y) = \mathcal{P}_0(|\mathbf{r}|/\sigma)^{|m|}e^{-\mathbf{r}^2/\sigma^2+im\phi}$ describes resonant pump with amplitude $\mathcal{P}_0$ and width $\sigma$, that can carry topological charge $m$ (the case of $m=0$, $\sigma\to\infty$ corresponds to uniform vortex-free pump). Setting $\ell = 2\ \mu\mathrm{m}$, $m_{\mathrm{eff}} = 5\times10^{-5}m_e$, one obtains the characteristic energy $\varepsilon_0 \approx 0.38\ \mathrm{meV}$ and time $\hbar\varepsilon_0^{-1} \approx 1.7\ \mathrm{ps}$. Further we set $p=4$, $w_0 = 0.5$, and $\alpha = 0.01$ - the parameters that can be attained in real experiments with polaritonic microcavities with high quality factors. Our results remain qualitatively similar in deeper potentials and for wider potential wells.

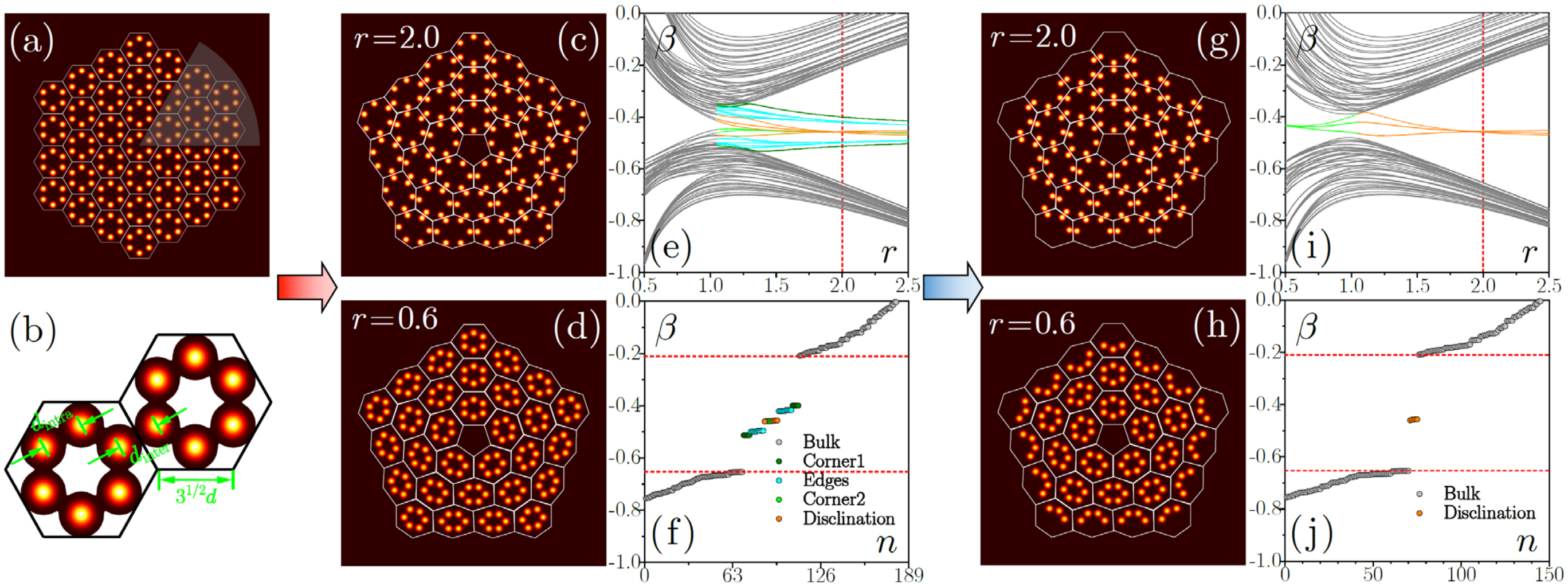

Fig. 1. (a) Original honeycomb array with $\mathcal{C}_6$ symmetry that is used for construction of arrays with disclination with indication of removed sector. (b) Schematic illustration of two cells of the array with indication of intra-pillar $d_{\mathrm{intra}}$ and inter-pillar $d_{\mathrm{inter}}$ spacings. Arrays with disclination with $\mathcal{C}_5$ symmetry obtained for different shifts of pillars corresponding to $r=2.0$ (c) and $r=0.6$ (d). (e) Eigenenergies of all linear modes of disclination array versus $r$. (f) Spectrum of the array of micropillars at $r=2.0$ corresponding to red vertical dashed line in (e). (g)-(j) show arrays with removed outer pillars, the dependence of eigenenergies of all their linear modes on $r$, and spectrum at $r=2.0$. Here and in all following figures we set the depth of potential wells $p=4$, their half-width $w_0=0.5$, original spacing between wells in honeycomb array $d=2.4$, and linear losses $\alpha=0.01$.

To construct a topological microcavity pillar array with disclination, we employ a two-step procedure. We start with finite honeycomb array [see Fig. 1(a) showing initial $\mathcal{R}(x,y)$ profile] with uniform spacing $d=2.4$ between pillars in the entire structure. We then introduce the Kekulé distortion into unit cells by shifting pillars along the direction perpendicular to the unit cell borders. The distortion is characterized by the shift parameter $r = d_{\mathrm{intra}} / d_{\mathrm{inter}}$ with $d_{\mathrm{intra}}$ and $d_{\mathrm{inter}}$ being intra- and inter-cell pillar spacings after shift [Fig. 1(b)]. In the absence of deformation $(r=1)$, the structure is periodic with $d_{\mathrm{intra}} = d_{\mathrm{inter}} = d$. At the next step we remove an angular sector of $2\pi/6$ from the structure and adjust the positions of remaining pillars such as to fill the removed sector, that leads to formation of disclination array with new $\mathcal{C}_5$ discrete rotational symmetry, not attainable with periodic arrays [Fig. 1(c),(d)]. Inclusion of a new sector and adjustment of pillar positions for its accommodation would instead create $\mathcal{C}_7$ array. Notice that this procedure creates disclination core in the array center. Increase of shift parameter from $r<1$ to $r>1$ values is accompanied by the shift of the Wannier centers [28,29] from topologically trivial positions in the center of each cell to nontrivial positions at the borders of unit cell, so that truncation may affect some of these centers at $r>1$. Thus, evaluating spectral charges at different unit cells of array with disclination, one can conclude that at $r<1$ (in trivial phase) the charge remains integer and equal to 1 for all cells, while at $r > 1$ (in topological phase) bulk cells still have unit charge, while cells at the disclination core and outer corner cells acquire fractional spectral charges. This is an indication of "filling anomaly" and the possibility of formation at the disclination core and in the outer corners of the array of localized states of topological origin at $r > 1$.

To illustrate this, we omit nonlinearity, pump and losses in Eq. (1) and calculate linear eigenmodes of $\mathcal{C}_5$ disclination array using the ansatz $\psi(x,y,t) = w(x,y)e^{-i\beta t}$, that leads to linear eigenvalue problem $\beta w = -(1/2)\Delta w - \mathcal{R} w$. This linear eigenvalue problem was solved using direct discretization of Laplacian operator in real space, with standard eigenvalue solver. The eigenmodes were obtained on the window $x, y \in [-60, +60]$ on $1201 \times 1201$ grid, i.e. for steps $dx = dy = 0.05$. Linear eigenenergies $\beta$ of all modes of the array (that are purely real in this case in contrast to disclination lattices with on-site gain and losses [34]) are shown in Fig. 1(e) vs shift parameter $r$, while details of linear spectrum at $r = 2$ are given in Fig. 1(f). One can see that at $r > 1$ topological gap opens in the spectrum (the region between gray bulk bands), where states localized at the disclination core (orange lines/ dots) coexist with corner and edge modes at the outer edge of the array (blue and green lines) (Fig. S1 in Supplemental Materials [64] provides examples of such localized modes). We stress that due to their topological nature the disclination states persist in spectrum even in the presence of moderate diagonal or off-diagonal disorder, provided that topological gap remains open. Since such states reside in the spectral gap, they are protected against hybridization with bulk states, allowing robust excitation with resonant pump (see below) as long as the resonance conditions are met.

## 3. Uniform pump

In this work we show that resonant pump $\mathcal{P}(x,y)e^{-i\varepsilon t}$ can be used for selective excitation of topological disclination states, when pump energy detuning $\varepsilon$ matches their eigenenergy $\beta$ in the gap. To avoid simultaneous excitation of disclination states and edge states with close eigenenergies at the outer boundary of the structure [Fig. 1(f)], we intentionally removed outer pillars, as shown in Fig. 1(g), (h), that drastically simplifies linear spectrum leaving in the gap only disclination states [orange lines in Fig. 1(i),(j)]. In Fig. 2(a) we show the dependence of peak amplitude $a = \max|w|$ of the disclination state excited by uniform pump $\mathcal{P}(x,y) = \mathcal{P}_0$ with $m = 0$ covering the entire array [Fig. 2(c)] versus detuning $\varepsilon$. At this point we take into account nonlinearity of the system. The above mentioned states excited by resonant pump are obtained in the form $\psi = we^{-i\varepsilon t}$ (where $e^{-i\varepsilon t}$ time dependence is dictated by the pump) from Eq. (1) using Newton iteration method. Newton iterations were performed on the window $x, y \in [-60, +60]$ on $1201 \times 1201$ grid, the iterations were continued until the tolerance of $10^{-14}$ (characterizing the difference between solutions on two subsequent iterations) is reached. Such pump has largest overlap with linear mode $w_{n=71}$ with 5 in-phase spots (here $n$ is the linear mode index, see Fig. S1 and discussion in [64]), whose eigenenergy is shown by orange dashed line in Fig. 2(a), and at very low pump amplitudes $\mathcal{P}_0$ the tip of the resonance curve coincides with eigenvalue $\varepsilon = \beta_{n=71}$ of this mode. Corresponding nonlinear state is localized at the disclination core [Fig. 2(h)], even though some background is present inside the array. As the pump amplitude increases, the resonance curve acquires a pronounced nonlinearity-induced tilt leading eventually to bistability [Fig. 2(a)]. Figures 2(d)-2(g) show that shape of the nonlinear disclination states strongly changes with pump detuning $\varepsilon$ and can be controlled by it. Good localization is achieved at the upper branch of the resonance curve at $\varepsilon > \beta_{n=71}$ with tendency for slight broadening near the tip. For too large $\mathcal{P}_0$ values the tip of the curve may shift into the band, where coupling with bulk modes causes sharp delocalization. The insets with phase distributions indicate the presence of internal currents in all nonlinear modes.

We present here the results for single value of linear loss coefficient $\alpha = 0.01$, but we checked that the structure of resonance curves remains qualitatively unchanged for larger background losses. As $\alpha$ increases, the width of the resonance curve increases, while maximal amplitude achieved in resonance and overall tilt of the curve decrease.

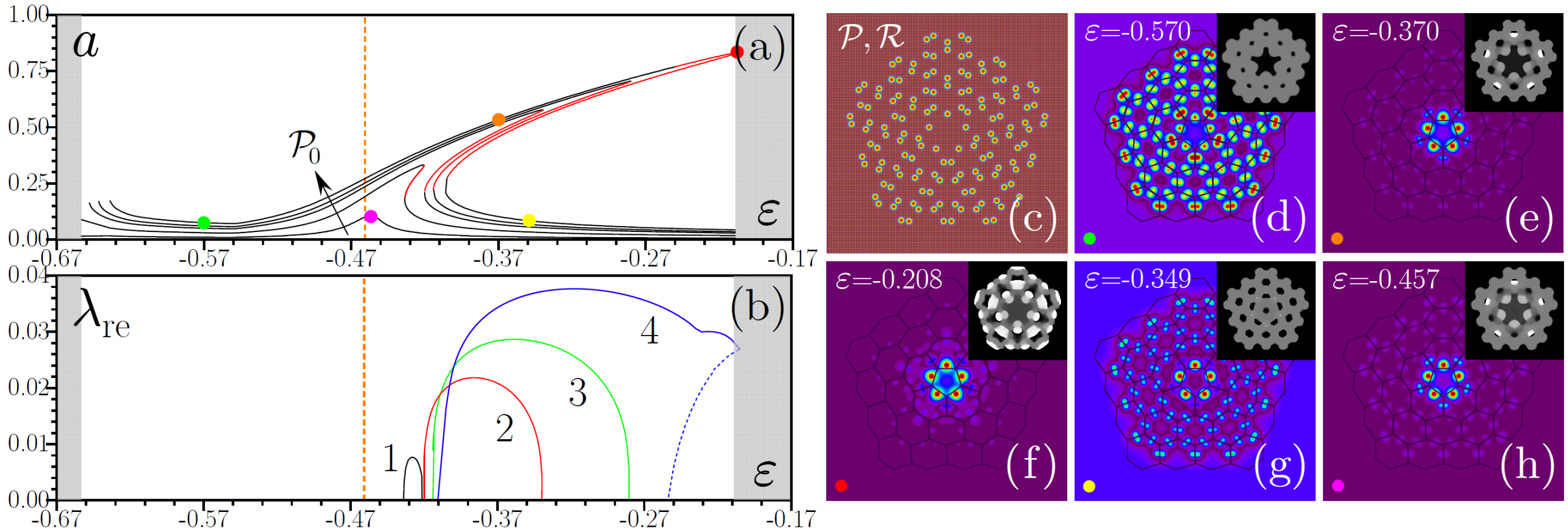

Fig. 2. (a) Peak amplitude $a = \max|w|$ of nonlinear disclination state vs detuning $\varepsilon$ for increasing amplitudes $\mathcal{P}_0 = 0.0005$, $0.0015$, $0.0025$, $0.0030$, and $0.0037$ of $m = 0$ pump, as indicated by the arrow. Grey regions indicate bands. Black lines – stable branches, red lines – unstable branches. (b) Perturbation growth rate vs $\varepsilon$ at $\mathcal{P}_0 = 0.0015$ (1), $0.0025$ (2), $0.0030$ (3), and $0.0037$ (4). (c) Profile of the array $\mathcal{R}$ with superimposed pump profile $\mathcal{P}$. (d) Modulus of polariton wavefunction in disclination states and its phase distribution (insets) at $\mathcal{P}_0 = 0.0037$ (d)-(g) and at $\mathcal{P}_0 = 0.0005$ (h) corresponding to the dots in (a). Lowest (black) level in color map of phase distributions corresponds to $-\pi$, while highest (white) level corresponds to $+\pi$. In all cases $r = 2.0$. Orange dashed lines in (a),(b) correspond to $\varepsilon = \beta_{n=71}$.

## 4. Stability analysis

To illustrate stability of nonlinear disclination states we use linear stability analysis by writing perturbed disclination states in the form $\psi(x,y,t)=[w(x,y)+u(x,y)e^{\lambda t}+v^*(x,y)e^{\lambda^* t}]e^{-i\varepsilon t}$, where $u,v\ll w$, and $\lambda=\lambda_{\rm re}+i\lambda_{\rm im}$ is the complex perturbation growth rate. Linearization of Eq. (1) around $w$ yields in this case the eigenproblem:

$$\begin{aligned}\lambda u&=i[+(1/2)\Delta u+\mathcal{R}u+i\alpha u+\varepsilon u-2|w|^2u-w^2v],\\ \lambda v&=i[-(1/2)\Delta v-\mathcal{R}v+i\alpha v-\varepsilon v+2|w|^2v+w^{*2}u].\end{aligned}\tag{2}$$

The state $w$ is stable when $\lambda_{\rm re}\le 0$ for all possible perturbations and is unstable when $\lambda_{\rm re}>0$ at least for one of them. The eigenvalue problem (2) was also solved using discretization in real space, with standard eigenvalue solver. Solution of (2) confirms stability of lower and upper branches of the resonance curves (black lines) and instability of the middle branches (red lines), even when resonances exhibit significant tilt [Fig. 2(a)]. Only for the largest pump amplitude, when resonance curve penetrates into band a small part of the upper branch near the tip becomes unstable. Representative $\lambda_{\rm re}(\varepsilon)$ dependencies are shown in Fig. 2(b), where solid lines correspond to middle branches and dashed one – to upper branch.

## 5. Vortical pump and evolution dynamics

One of the most interesting features of the array with disclination is the presence in its spectrum in topological gap at $r>1$ of degenerate pairs of eigenmodes (two pairs in $\mathcal{C}_5$ case), whose linear combinations can produce $m=\pm 1$ or $\pm 2$ *vortex disclination states*. Vortex-carrying pump overlaps with and can efficiently excite namely these vortex states, illustrating a new degree of freedom for control of excitations in polaritonic HOTIs with disclination. We thus use the pump profile $\mathcal{P}(x,y)=\mathcal{P}_0(|\mathbf{r}|/\sigma)^{|m|}e^{-\mathbf{r}^2/\sigma^2+im\phi}$ with $m\ne 0$ and adjust the pump width $\sigma$ such that it efficiently covers pillars at the disclination core [see Fig. 3(b),(f),(j)]. Notice that for larger pump widths the efficiency of excitation of modes at the disclination core gradually decreases and that of the bulk modes increases, but resonances with bulk modes (occurring in gray bulk bands, not shown here) typically remain broad and have lower amplitudes in comparison with in-gap resonance. It turns out that in polariton HOTI with disclination, the impact of resonant pump on condensate is fundamentally influenced by the system's topology. Thus, in array with shift parameter $r=2$ the $m=1$ pump efficiently excites the combination $w_{n=72}+iw_{n=73}$ of disclination modes, while repulsive nonlinearity tilts the resonance curve [Fig. 3(a)] and allows to considerably tune the profile of the nonlinear vortex disclination state [see Fig. 3(c),(d) and insets with phase distributions]. Similar picture is observed for $m=2$ pump exciting nonlinear vortex state originating from combination $w_{n=74}+iw_{n=75}$ of linear modes [see resonance curves in Fig. 3(e) and typical profiles in Fig. 3(g),(h)]. Importantly, linear stability analysis reveals stability of vortex states from upper and lower branches in Fig. 3(a) and 3(e) (see black lines) for *both* $m=1$ and $m=2$ charges, in complete contrast to nonlinear vortex states in conservative HOTIs with disclinations, where one of these states is always unstable [36]. It should be mentioned that when amplitude $\mathcal{P}_0$ of vortical pump becomes sufficiently large, the tip of the resonance curve may shift into the band. Corresponding nonlinear disclination states couple with bulk modes and delocalize. The onset of such delocalization is visible already for states corresponding to the red and blue points in Fig. 3(a) and 3(e) that acquire noticeable background in the entire array. Delocalization of such states implies that they also loose topological protection because nonlinearity shifts them away from topological gap.

Moreover, an intriguing result is that pump beams with vorticity $m=3,4,\ldots$ can excite nonlinear disclination states too. However, as the modes with such charges are prohibited by the discrete rotational symmetry of the $\mathcal{C}_5$ system [see [65,66], where it is shown that only vortex states with $|m|\le(k-1)/2$ are allowed in $\mathcal{C}_k$ system with odd $k$], the latter finds rather elegant way to overcome this restriction through the formation of vortex clusters. Thus, for $m=3$ pump one observes the formation of disclination state carrying $m=-2$ phase singularity in the center surrounded by five $m=+1$ off-center singularities [see Fig. 3(i)-3(l)]. As linear stability analysis shows (see Fig. S2 in [64] for corresponding $\lambda_{\rm re}(\varepsilon)$ dependencies), this state can be stable too for a broad range of pump amplitudes $\mathcal{P}_0$. Similarly, pump with $m=4$ produces nonlinear states with $m=-1$ singularity in the center surrounded by five $m=+1$ singularities, etc. Thus, while the arrangement of singularities is dictated by the discrete rotational symmetry of the system, the presence of the topological gap ensures localization of polaritons with in-gap energies at the disclination core of the structure.

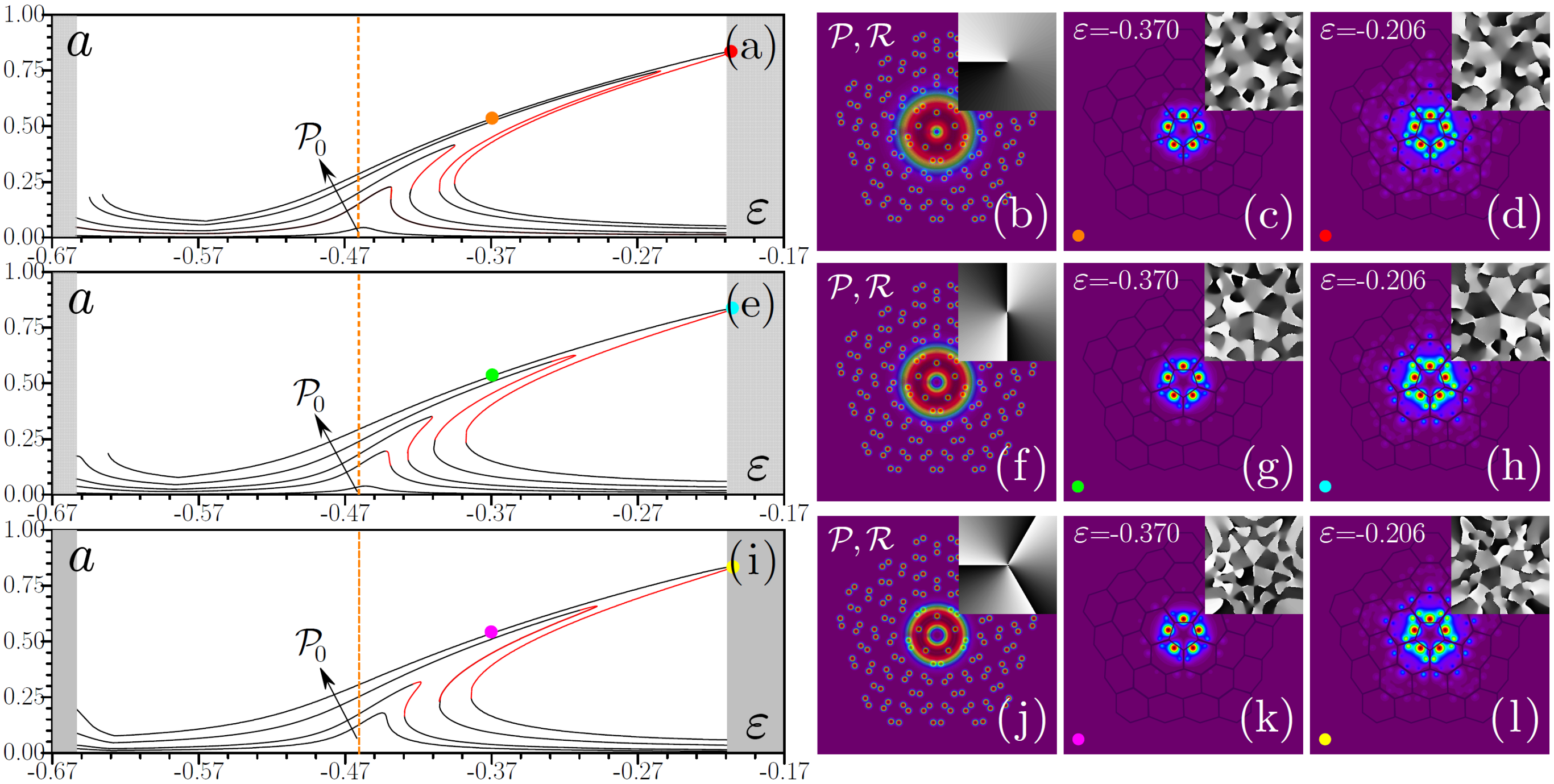

Fig. 3. (a) Peak amplitude $a = \max|w|$ of vortex disclination states versus detuning $\varepsilon$ for progressively increasing pump amplitudes (a) from $\mathcal{P}_0 = 0.0005$ to $\mathcal{P}_0 = 0.0110$ for $m = 1$, (e) from $\mathcal{P}_0 = 0.0005$ to $\mathcal{P}_0 = 0.0145$ for $m = 2$, and (i) from $\mathcal{P}_0 = 0.0025$ to $\mathcal{P}_0 = 0.0175$ for $m = 3$, as shown by the arrows. Examples of vortex pump profiles $\mathcal{P}$ superimposed on $\mathcal{R}$ landscape and vortex disclination states corresponding to dots of different color for $m = 1$ (b)-(d), $m = 2$ (f)-(h), and $m = 3$ (j)-(l) pump. Lowest (black) level in color map of phase distributions corresponds to $-\pi$, while highest (white) level corresponds to $+\pi$. The pump width $\sigma = 5$ (b), $\sigma = 4$ (f), and $\sigma = 3$ (j), while $r = 2.0$. Orange dashed lines correspond to $\varepsilon = \beta_{n=72}$ (a) and $\varepsilon = \beta_{n=74}$ (e),(i).

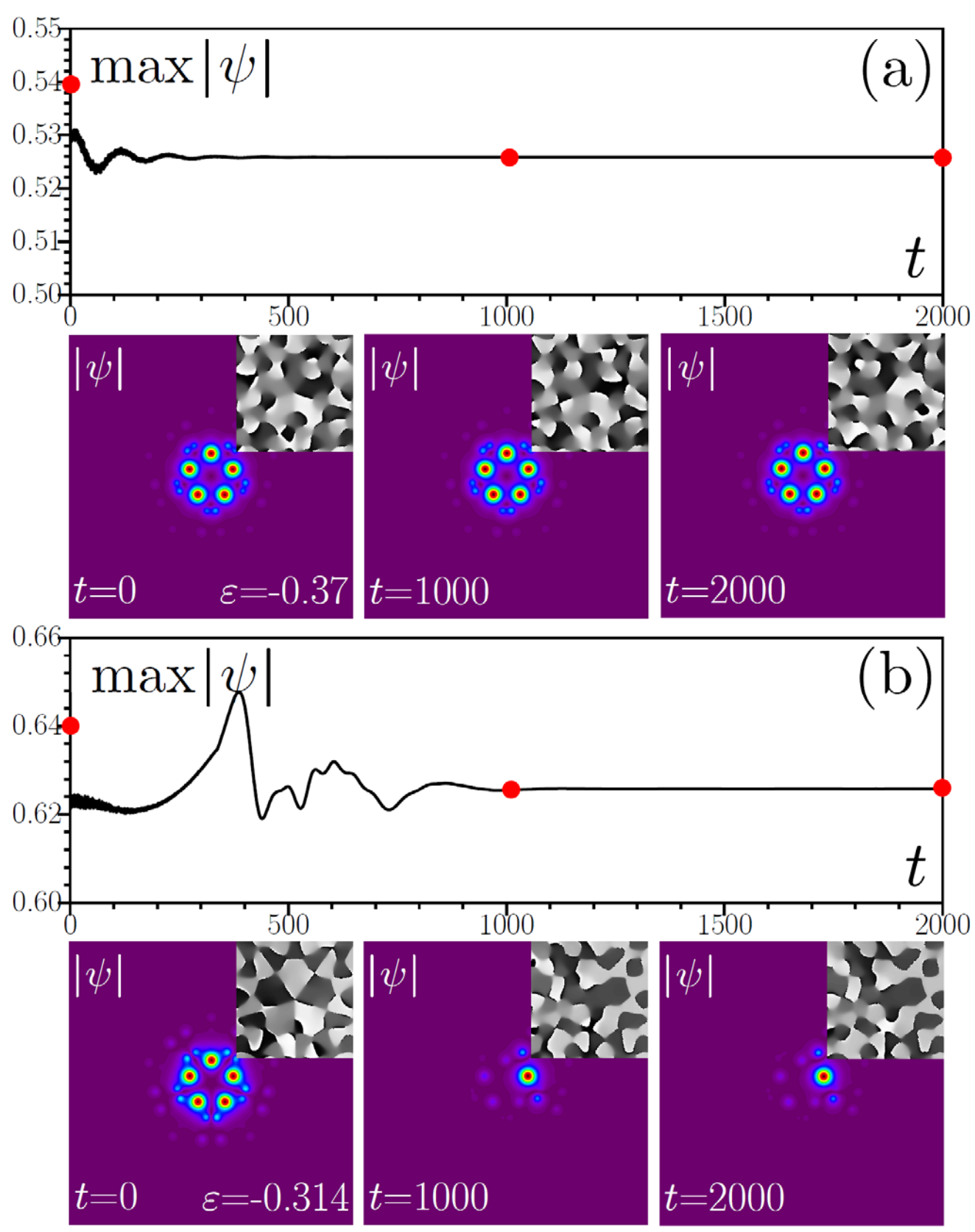

Fig. 4. Peak amplitude $\max|\psi|$ of the state vs time illustrating stable evolution of perturbed disclination state with $m = 1$ at $\varepsilon = -0.370$, $\mathcal{P}_0 = 0.0110$ (a), and decay of the unstable state with $m = 2$ at $\varepsilon = -0.314$, $\mathcal{P}_0 = 0.0145$ (b). Modulus and phase distributions (insets) correspond to red dots in $\max|\psi(t)|$ dependencies.

Finally, to analyze the dynamics of disclination states with different topological charges $m$ beyond the perturbative approach, we modeled evolution of such states in the frames of Eq. (1). This equation was solved using split-step fast Fourier transform method. The nonlinear and pump steps in the frames of this method were calculated using fourth-order Runge-Kutta method with typical time step $dt = 10^{-3}$. The input states were perturbed at $t = 0$ by the broadband complex noise $\xi(x, y)$ (up to $5\%$ in amplitude). Figure 4(a) illustrates typical stable evolution of $m = 1$ vortex disclination state from the upper branch for pump amplitude $\mathcal{P}_0 = 0.0110$. In this case the perturbed disclination state "cleans up" from the noise at typical times $\sim 5\alpha^{-1}$ that is consistent with results of the linear stability analysis and then keeps its structure with time as long as pump is on. The example of unstable evolution of the $m = 2$ vortex state from upper branch that transforms into asymmetric state at $\mathcal{P}_0 = 0.0085$ is shown in Fig. 4(b) (for evolution of $m = 0$ state see Fig. S3 in [64]).

## 6. Summary

Summarizing, we have shown that polariton HOTIs with disclination support a rich variety of stable nonlinear disclination states, including vortex-carrying ones, that can be resonantly excited with structured pump beams with energy detuning in topological gap of the system. Among the advantages of this system is the possibility to control the localization degree and structure of disclination modes by simple variation of detuning of the pump. Our analysis has revealed drastic differences in stability properties of vortex disclination states in this system in comparison with conservative nonlinear structures with disclinations. Our results advance the understanding of higher-order topological phases in aperiodic driven-dissipative systems and may be relevant for other areas of physics, including design of topological lasers on aperiodic lattices.

## Acknowledgements

This work was supported by the Russian Science Foundation (grant 24-12-00167) and partially by the research project FFUU-2024-0003 of the Institute of Spectroscopy of the Russian Academy of Sciences.